\documentclass[twoside,twocolumn,9pt]{article}
\usepackage{extsizes}
\usepackage[super,sort&compress,comma]{natbib} 
\usepackage[version=3]{mhchem}
\usepackage[left=1.5cm, right=1.5cm, top=1.785cm, bottom=2.0cm]{geometry}
\usepackage{balance}
\usepackage{widetext}
\usepackage{times,mathptmx}
\usepackage{sectsty}
\usepackage{graphicx} 
\usepackage{lastpage}
\usepackage[format=plain,justification=raggedright,singlelinecheck=false,font={stretch=1.125,small,sf},labelfont=bf,labelsep=space]{caption}
\usepackage{float}
\usepackage{fancyhdr}
\usepackage{fnpos}
\usepackage[english]{babel}
\usepackage{array}
\usepackage{droidsans}
\usepackage{charter}
\usepackage[T1]{fontenc}
\usepackage[usenames,dvipsnames]{xcolor}
\usepackage{setspace}
\usepackage[compact]{titlesec}
\usepackage[english]{babel}
\usepackage[utf8]{inputenc}
\usepackage{mathtools}
\usepackage{graphicx}
\usepackage{nicefrac}
\usepackage{url}
\usepackage[normalem]{ulem}

\usepackage{color}

\newcommand*{\mat}[1]{\mathbf{#1}}
\newcommand*{\coord}[1]{\mathbf{#1}}
\newcommand*{\vecr}{\coord{r}}
\newcommand*{\vecx}{\coord{x}}

\newcommand*{\binteg}[3]{\int^{\mathrlap{#3}}_{\mathrlap{#2}}\ud{#1}\,}
\newcommand*{\integ}[1]{\!\int\!\ud{#1}\:}
\newcommand*{\iinteg}[2]{\integ{#1}\!\!\integ{#2}}

\DeclareMathOperator{\erf}{erf}
\DeclareMathOperator{\Trace}{\mathrm{Tr}}

\DeclarePairedDelimiter{\abs}{\lvert}{\rvert}
\DeclarePairedDelimiterX\braket[2]{\langle}{\rangle}{#1\delimsize\vert#2}

\newcommand*{\du}{\partial}
\newcommand*{\e}{\mathrm{e}}
\newcommand*{\eqspace}{\phantom{{} = {}}}
\newcommand*{\half}{\frac{1}{2}}
\newcommand*{\isDefinedAs}{\coloneqq}
\newcommand*{\nhalf}{\nicefrac{1}{2}}
\newcommand*{\ud}{\mathrm{d}}

\allowdisplaybreaks[4]

\definecolor{cream}{RGB}{222,217,201}

\begin{document}

\pagestyle{fancy}
\thispagestyle{plain}
\fancypagestyle{plain}{

\fancyhead[C]{\includegraphics[width=18.5cm]{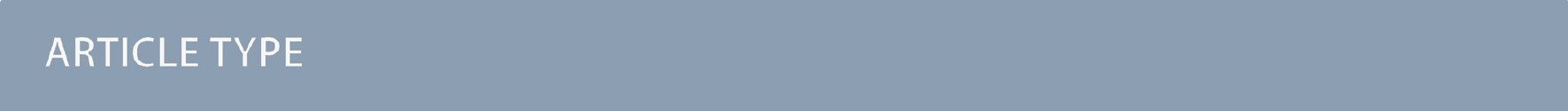}}
\fancyhead[L]{\hspace{0cm}\vspace{1.5cm}\includegraphics[height=30pt]{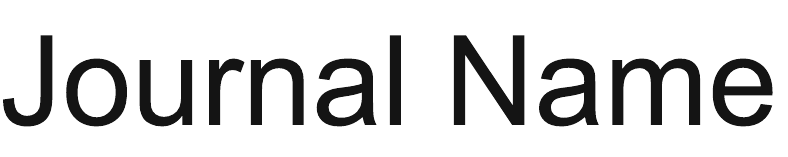}}
\fancyhead[R]{\hspace{0cm}\vspace{1.7cm}\includegraphics[height=55pt]{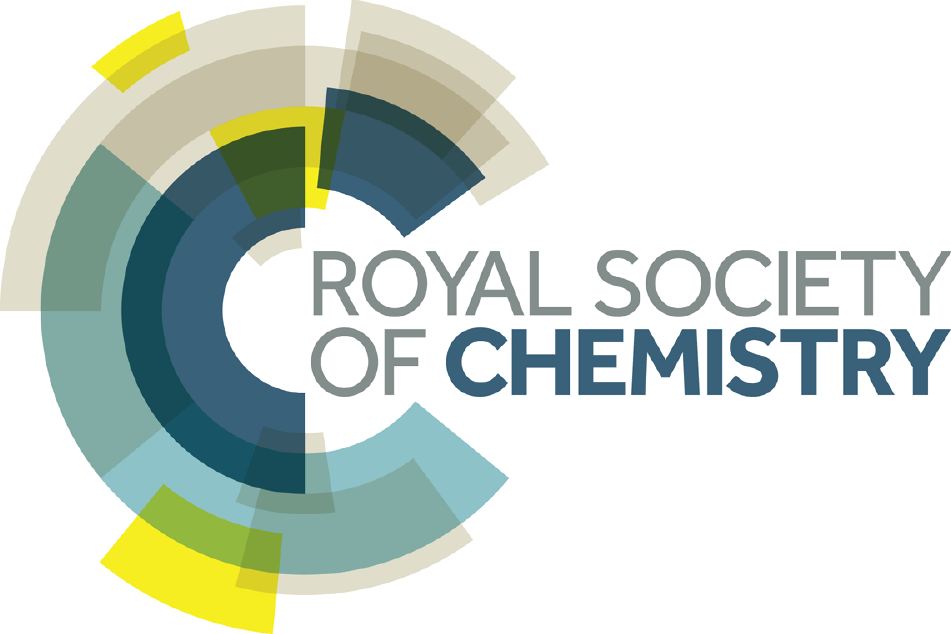}}
\renewcommand{\headrulewidth}{0pt}
}

\makeFNbottom
\makeatletter
\renewcommand\LARGE{\@setfontsize\LARGE{15pt}{17}}
\renewcommand\Large{\@setfontsize\Large{12pt}{14}}
\renewcommand\large{\@setfontsize\large{10pt}{12}}
\renewcommand\footnotesize{\@setfontsize\footnotesize{7pt}{10}}
\makeatother

\renewcommand{\thefootnote}{\fnsymbol{footnote}}
\renewcommand\footnoterule{\vspace*{1pt}%
\color{cream}\hrule width 3.5in height 0.4pt \color{black}\vspace*{5pt}} 
\setcounter{secnumdepth}{5}

\makeatletter 
\renewcommand\@biblabel[1]{#1}            
\renewcommand\@makefntext[1]%
{\noindent\makebox[0pt][r]{\@thefnmark\,}#1}
\makeatother 
\renewcommand{\figurename}{\small{Fig.}~}
\sectionfont{\sffamily\Large}
\subsectionfont{\normalsize}
\subsubsectionfont{\bf}
\setstretch{1.125} 
\setlength{\skip\footins}{0.8cm}
\setlength{\footnotesep}{0.25cm}
\setlength{\jot}{10pt}
\titlespacing*{\section}{0pt}{4pt}{4pt}
\titlespacing*{\subsection}{0pt}{15pt}{1pt}

\fancyfoot{}
\fancyfoot[LO,RE]{\vspace{-7.1pt}\includegraphics[height=9pt]{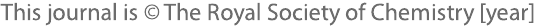}}
\fancyfoot[CO]{\vspace{-7.1pt}\hspace{13.2cm}\includegraphics{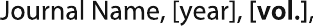}}
\fancyfoot[CE]{\vspace{-7.2pt}\hspace{-14.2cm}\includegraphics{RF}}
\fancyfoot[RO]{\footnotesize{\sffamily{1--\pageref{LastPage} ~\textbar  \hspace{2pt}\thepage}}}
\fancyfoot[LE]{\footnotesize{\sffamily{\thepage~\textbar\hspace{3.45cm} 1--\pageref{LastPage}}}}
\fancyhead{}
\renewcommand{\headrulewidth}{0pt} 
\renewcommand{\footrulewidth}{0pt}
\setlength{\arrayrulewidth}{1pt}
\setlength{\columnsep}{6.5mm}
\setlength\bibsep{1pt}

\makeatletter 
\newlength{\figrulesep} 
\setlength{\figrulesep}{0.5\textfloatsep} 

\newcommand{\topfigrule}{\vspace*{-1pt}%
\noindent{\color{cream}\rule[-\figrulesep]{\columnwidth}{1.5pt}} }

\newcommand{\botfigrule}{\vspace*{-2pt}%
\noindent{\color{cream}\rule[\figrulesep]{\columnwidth}{1.5pt}} }

\newcommand{\dblfigrule}{\vspace*{-1pt}%
\noindent{\color{cream}\rule[-\figrulesep]{\textwidth}{1.5pt}} }

\makeatother

\twocolumn[
  \begin{@twocolumnfalse}
\vspace{3cm}
\sffamily
\begin{tabular}{m{4.5cm} p{13.5cm} }

\includegraphics{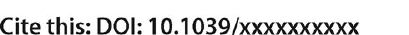} & \noindent\LARGE{\textbf{Avoiding the 4-index transformation in one-body reduced density matrix functional calculations for separable functionals}} \\
\vspace{0.3cm} & \vspace{0.3cm} \\

 & \noindent\large{Klaas J.H. Giesbertz,$^{\ast}$\textit{$^{a}$}} \\

\includegraphics{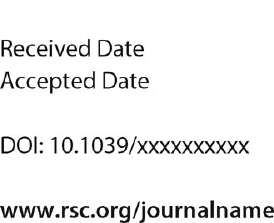} & \noindent\normalsize{One of the major computational bottlenecks in one-body reduced density matrix (1RDM) functional theory is the evaluation of approximate 1RDM functionals and their derivatives. The reason is that more advanced approximate functionals are almost exclusively defined in the natural orbital basis, so a 4-index transformation of the two-electron integrals appears to be unavoidable. I will show that this is not the case and that so-called separable functionals can be evaluated much more efficiently, i.e.\ only at cubic cost in the basis size. Since most approximate functionals are actually separable, this new algorithm is an important development to make 1RDM functional theory calculations feasible for large electronic systems.} \\

\end{tabular}

 \end{@twocolumnfalse} \vspace{0.6cm}

  ]

\renewcommand*\rmdefault{bch}\normalfont\upshape
\rmfamily
\section*{}
\vspace{-1cm}


\footnotetext{\textit{$^{\ast}$~Section of Theoretical Chemistry, Faculty of Exact Sciences, VU University, De Boelelaan 1083, 1081 HV Amsterdam, The Netherlands. Tel: XX XXXX XXXX; E-mail: k.j.h.giesbertz@vu.nl}}

%



\section{Introduction}
\label{sec:intro}
Though density functional theory (DFT) is formally exact, practical density functionals have great difficulty to capture strongly correlated phenomena such as the breaking of chemical bonds.\cite{PerdewParrLevy1982, OssowskiBoyerMehl2003, RuzsinszkyPerdewCsonka2006} The calculation of excitation energies along the bond-breaking coordinate with the current approximations results in an even bigger disaster.\cite{MaitraTempel2006, GiesbertzBaerends2008, FuksRubioMaitra2011} Also long-range charge transfer excitations pose a serious challenge for semi-local density functionals,\cite{CasidaGutierrezGuan2000, DreuwWeismanHead-Gordon2003} though some improvements have been reported with the help of range-separated hybrids,\cite{YanaiTewHandy2004, PeachBenfieldHelgaker2008} direct modifications of the kernel\cite{CasidaGutierrezGuan2000, GritsenkoBaerends2004, NeugebauerGritsenkoBaerends2006} or the variational approach by Ziegler et al.\cite{ZieglerSethKrykunov2008, ZieglerSethKrykunov2009, ZieglerKrykunov2010}

One-body reduced density matrix (1RDM)functional theory provides a promising route to alleviate most of these problems existent in practical DFT. It has been demonstrated that the ground state energy of small singly bonded molecular systems can be reasonably well reproduced along the full bond-breaking coordinate.\cite{GritsenkoPernalBaerends2005, RohrPernalGritsenko2008, PirisMatxainLopez2010, MentelMeerGritsenko2014, Piris2014} It has been shown for two-electron systems that the time-dependent extension of 1RDM functional is much more capable in dealing with bond breaking excitations and charge transfer excitations even within the adiabatic approximation, and also a significant amount of double excitations is captured.\cite{GiesbertzBaerendsGritsenko2008, GiesbertzPernalGritsenko2009, GiesbertzGritsenkoBaerends2010a, GiesbertzGritsenkoBaerends2010b} Attempts are currently made to extend these results to general $N$-electron systems.\cite{Pernal2012, ChatterjeePernal2012, MeerGritsenkoBaerends2014} A more extensive overview of the current status of 1RDM functional theory can be found in Ref.~\citenum{PernalGiesbertz2015}.

Though 1RDM functional theory has some appealing advantages compared to DFT, its practical use is currently limited due to two major computational bottle necks. One computational hurdle is the excruciatingly slow self-consistent field (SCF) convergence to obtain the ground state energy. Although several algorithms have been proposed,\cite{CioslowskiPernal2001, CohenBaerends2002, Pernal2005, PirisUgalde2009} a significant breakthrough on this difficulty has not yet been achieved. Another computational complication is the evaluation of the 1RDM functionals themselves. Only the simplest approximate 1RDM functionals are explicitly defined in terms of the 1RDM. More advanced approximations are defined implicitly via the natural orbitals (NOs) and (natural) occupation numbers, which are defined as the eigenfunctions and eigenvalues of the 1RDM respectively
\begin{align}
\gamma(\vecx,\vecx') = \sum_kn_k\,\phi_k(\vecx)\phi^*_k(\vecx') \, ,
\end{align}
where $\vecx \isDefinedAs \vecr\sigma$ is a combined space-spin coordinate.
Current implementations therefore rely on a 4-index transformation of the two-electron integrals to the NO basis which is a very costly operation and impairs any calculation on systems with a large number of electrons. The situation is far worse than in correlated methods such as coupled cluster (CC) and configuration interaction (CI), since the 4-index transformation needs to be performed at each step of the SCF procedure.

I will show in this article that the 4-index transformation can actually be avoided for so-called separable functionals. Separability allows a functional to evaluated directly in the atomic orbital (AO) basis, or any other basis employed in the computer code. This reduces the computational cost from formal $m^5$ to a formal $m^4$ scaling, where $m$ is the size of the basis set. Most integral routines make use of screening techniques to only calculate significant two-electron integrals which further reduces the scaling to $m^3$ or even less. It turns out that most current approximate 1RDM functionals are separable functionals, with a only a few exceptions.

The paper is organized as follows. First the algorithm for the evaluation of separable 1RDM functionals and the first order derivatives is explained in detail. Particular attention is needed for the so-called diagonal corrections which are sometimes called `self-interaction corrections'. Special care needs to be taken to avoid excessive computational cost and memory imprint for these corrections. The evaluation of some particular functionals will be discussed to illustrate the algorithm. The next section presents benchmark results for the new algorithm using alkanes of varying length as test systems. A significant speed-up is obtained for the evaluation of separable functionals and is most drastic for functionals without diagonal corrections.

\section{Algorithm}
\label{sec:algo}
To avoid the 4-index transformation we generalize the direct approach in Hartree--Fock (HF) theory to calculate the contribution of the two-body interaction to the energy and the Fock matrix.\cite{AlmlofFaegriKorsell1982} For example, consider the Hartree (classical Coulomb) contribution to the energy which in terms of the HF orbitals has the following simple form
\begin{align}\label{eq:Hartree}
W^{\text{H}} = \half\sum_{ij}n_in_j \left[ ii | jj \right] \, ,
\end{align}
where $n_i$ are the occupation numbers of the HF orbitals, i.e.\ the NOs of the HF system. The two-electron integrals are denoted in the chemist's notation, so they are defined as
\begin{multline}
\left[ ij | kl \right] \isDefinedAs
\iinteg{\vecx_1}{\vecx_2}\phi^*_i(\vecx_1)\phi_j(\vecx_1)
w(\abs{\vecr_1 - \vecr_2}) \phi^*_k(\vecx_2)\phi_l(\vecx_2) \, .
\end{multline}
The interaction between the particles will be usually the Coulomb interaction $w(r) = 1/r$, but could also have some other form, e.g.\ the long-range part of the Coulomb interaction in a range-separated scheme $w(r) = \erf(\mu r) / r$.\cite{Savin1988, Savin1995, RohrToulousePernal2010}

In an arbitrary basis, $\{\chi_{\mu}\}$, the Hartree term can be expressed as
\begin{align}
W^{\text{H}} = \half\;\sum_{\mathclap{\mu\nu\kappa\lambda}}
\gamma_{\nu\mu}\gamma_{\lambda\kappa} \left[ \mu\nu | \kappa\lambda \right] \, ,
\end{align}
where the matrix elements of the 1RDM are defined as
\begin{align}
\gamma_{\mu\nu} \isDefinedAs \iinteg{\vecx}{\vecx'}\chi^*_{\mu}(\vecx)\gamma(\vecx,\vecx')\chi_{\nu}(\vecx') \, .
\end{align}
Throughout the text I will always use Greek indices to refer to this arbitrary basis in which the two-electron integrals are supplied. The latin indices will exclusively be used for the NO basis.

Suppose now that all the two-electron integrals are available in the basis $\{\chi_{\mu}\}$. Typically this will be an atomic orbital basis or a plane wave basis. The Hartree contribution can be calculated without transforming any 4-index quantity by first performing the contraction over only one 1RDM as
\begin{align}\label{eq:HartreePotential}
v^{\text{H}}_{\mu\nu} = \sum_{\kappa\lambda}\left[ \mu\nu | \kappa\lambda \right] \gamma_{\lambda\kappa} \, .
\end{align}
The summations are generally performed by looping over integrals stored on file or by calculating all integrals on the fly. In a second step the contraction with the other 1RDM is performed
\begin{align}
W^{\text{H}} = \half\Trace\Bigl\{\mat{\gamma}\cdot\mat{v}^{\text{H}}\Bigr\}
= \half\sum_{\mu\nu}\gamma_{\nu\mu} v^{\text{H}}_{\mu\nu} \, .
\end{align}
The only essential feature of the Hartree contribution to allow for this trick is that it is a Coulomb-type separable functional. With a Coulomb-type separable functional I mean a functional which can be expressed as a linear combination of a few terms of the form
\begin{align}\label{eq:CoulombSepar}
W^{\text{J}}(\mat{f},\mat{g})
= \half\;\sum_{\mathclap{\mu\nu\kappa\lambda}}f_{\nu\mu} \, g_{\lambda\kappa} \left[ \mu\nu | \kappa\lambda \right] \, .
\end{align}
This is a Coulomb-type separable functional which can efficiently be evaluated in the same fashion as the Hartree term.
Typically we will have $\mat{f} = \mat{g}$ and $\mat{f} = \mat{f}^{\dagger}$, but this is not necessary for the trick to work. Its evaluation proceeds again via a Coulomb-like potential~\eqref{eq:HartreePotential}, which is generalized now to
\begin{align}\label{eq:HPotential}
v^{\text{J}}_{\mu\nu}(\mat{g}) = \sum_{\kappa\lambda}\left[ \mu\nu | \kappa\lambda \right] g_{\lambda\kappa} \, .
\end{align}
Likewise, an exchange-type separable functional can be expressed as a linear combination of a few terms of the form
\begin{align}\label{eq:exchangeSepar}
W^{\text{K}}(\mat{f},\mat{g})
= \half\;\sum_{\mathclap{\mu\nu\kappa\lambda}}f_{\lambda\mu} \, g_{\nu\kappa} \left[ \mu\nu | \kappa\lambda \right] \, ,
\end{align}
so just two indices are swapped around compared to the Coulomb case.
The exchange-type separable functionals allow for a similar efficient evaluation as the Coulomb-type separable functionals by first forming an exchange-type potential
\begin{align}\label{eq:xPotential}
v^{\text{K}}_{\mu\nu}(\mat{g}) = \sum_{\kappa\lambda}g_{\lambda\kappa} \left[ \mu\lambda | \kappa\nu \right]
\end{align}
and subsequently performing the final contraction
\begin{align}
W^{\text{K}}(\mat{f},\mat{g}) &= \half\sum_{\mu\nu}f_{\nu\mu}v^{\text{K}}(\mat{g})_{\mu\nu}
= \half\Trace\Bigl\{\mat{f}\cdot\mat{v}^{\text{K}}(\mat{g})\Bigr\} \, .
\end{align}
When the basis functions are real, the Coulomb- and exchange-type versions are the only two possible distinct forms of a separable functional. In the case one uses complex basis functions, e.g.\ plane waves, one also might need to consider the versions where the complex conjugation has been interchanged in the last pair of the two-electron integral, i.e.\ $\left[ \mu\nu | \kappa\lambda \right] \to \left[ \mu\nu | \lambda\kappa \right]$ in~\eqref{eq:CoulombSepar} and~\eqref{eq:exchangeSepar}.

Not only the contribution to the energy can be calculated in this manner. Also the projected orbital derivatives
\begin{align}
W_{kl} \isDefinedAs \integ{\vecx}\frac{\du W}{\du\phi_k(\vecx)}\phi_l(\vecx) \, ,
\end{align}
needed for the construction of the Fock matrix\cite{Pernal2005, PirisUgalde2009} or direct orbital optimization,\cite{CioslowskiPernal2001} can easily be calculated. Only the final contraction over the potentials needs to be left out
\begin{align}
W^{\text{J/K}}_{\mu\nu}(\mat{f},\mat{g})
= \half\sum_{\kappa}
\Bigl(f_{\mu\kappa}v^{\text{J/K}}_{\kappa\nu}(\mat{g}) + g_{\mu\kappa}v^{\text{J/K}}_{\kappa\nu}(\mat{f})\Bigr)
\end{align}
and subsequently a transformation to the NO basis needs to be made. For most approximate functionals, $\mat{f}$ and $\mat{g}$ will be diagonal in the NO basis. So it is computationally beneficial to transform first the potentials to the NO basis and only then to multiply them by the diagonal $\mat{f}$ and $\mat{g}$ matrices
\begin{align}
W^{\text{J/K}}_{kl}(\mat{f},\mat{g})
= \half\Bigl(f_kv^{\text{J/K}}_{kl}(\mat{g}) + g_kv^{\text{J/K}}_{kl}(\mat{f})\Bigr) \, ,
\end{align}
where $f_k$ and $g_k$ denote the diagonal entries of the $\mat{f}$ and $\mat{g}$ matrices in the NO representation. The energetic contribution can now cheaply be obtained by taking the trace
\begin{align}
W^{\text{J/K}}(\mat{f},\mat{g}) = \half\Trace\Bigl\{\mat{W}^{\text{J/K}}(\mat{f},\mat{g})\Bigr\} \, .
\end{align}
If the matrices $\mat{f}$ and $\mat{g}$ are not only diagonal in the NO basis, but if also their diagonal elements only depend on the occupation numbers with the same index, i.e.\ $f_k(\mat{\gamma}) = f_k(n_k)$ and $g_k(\mat{\gamma}) = g_k(n_k)$, the derivatives with respect to the occupation numbers become particularly simple
\begin{align}
\frac{\du W^{\text{J/K}}}{\du n_k}
= \half\left(\frac{\du f_k}{\du  n_k}v^{\text{J/K}}_{kk}(\mat{g}) + \frac{\du g_k}{\du n_k}v^{\text{J/K}}_{kk}(\mat{f})\right) \, .
\end{align}
Most approximate 1RDM functionals exhibit such a simple dependence on the occupation numbers. More complicated dependencies need to be worked out for each approximate functional separately.

Though separability might seem to be a very stringent condition on a general 1RDM functional, many approximate functionals used in 1RDM functional calculations are actually separable. The only non-separable functionals I am aware of are the empirical functional by Marques and Lathiotakis\cite{MarquesLathiotakis2008}, the automated version of the BBC3 functional by Rohr et al.\cite{RohrPernalGritsenko2008} and the PNOF4 by Piris et al.\cite{PirisMatxainLopez2010} To use the proposed scheme to avoid the 4-index transformation, one only needs to rewrite the approximate 1RDM functional in a separable form, i.e.\ as a linear combination of terms of the form~\eqref{eq:CoulombSepar} and~\eqref{eq:exchangeSepar}. An obvious example is the Müller functional,\cite{Muller1984, PhD-Buijse1991, BuijseBaerends2002} since it was originally published in its separable form
\begin{align}\label{eq:Mueller}
W^{\text{Müller}}
&= \half\Trace\Bigl\{\mat{\gamma}\cdot\mat{v}^{\text{J}}(\mat{\gamma}) - 
\sqrt{\mat{\gamma}}\cdot\mat{v}^{\text{K}}(\sqrt{\mat{\gamma}})\Bigr\} \, .
\end{align}
A function of the 1RDM, the square root in this case, is defined in the usual manner via its diagonal representation
\begin{align}
f(\mat{\gamma})_{\mu\nu} \isDefinedAs \sum_i\braket{\chi_{\mu}}{\phi_i}\, f(n_i) \, \braket{\phi_i}{\chi_{\nu}} \, .
\end{align}
Approximate functionals which simply modify the square root to some other power, $\sqrt{\mat{\gamma}} \to \mat{\gamma}^{\alpha}$, also belong to this class of functionals which the 4-index transformation can trivially be avoided.\cite{Holas1999, CsanyiArias2000, SharmaDewhurstLathiotakis2008, LathiotakisSharmaDewhurst2009}

An explicit expression in terms of the 1RDM itself is not available for more advanced 1RDM functionals which classify NOs in strongly and weakly occupied groups and\slash{}or contain `diagonal corrections'. Though these functionals are only explicitly defined in terms of occupation numbers and NOs, many of these functionals can still be rewritten in a separable form, allowing for the previously described tricks. These non-trivial separable forms of more advanced 1RDM functionals are best explained with an example. To this end we will use the BBC2 functional, which is defined in the NO representation as\cite{GritsenkoPernalBaerends2005}
\begin{align}
W^{\text{BBC2}} \isDefinedAs W^{\text{H}} + \half\sum_{ij}F(n_i,n_j) \left[ ij | ji \right] \, ,
\end{align}
where $W^{\text{H}}$ is the usual Hartree term introduced earlier~\eqref{eq:Hartree} and $F(n_i,n_j)$ is defined as 
\begin{align}\label{eq:BBC2Fdef}
\raisetag{\baselineskip}
F(n_i,n_j) \isDefinedAs \begin{cases}
\sqrt{n_in_j}	&\text{for $i \neq j$ and $n_i,n_j < \nhalf$} \\
-n_in_j		&\text{for $i \neq j$ and $n_i,n_j \geq \nhalf$} \\
-\sqrt{n_in_j}	&\text{otherwise} \, .
\end{cases}
\end{align}
The involved expression for $F(n_i,n_j)$ renders an explicit expression in terms of $\mat{\gamma}$ virtually impossible. Nevertheless, the BBC2 functional can still be expressed in a separable form by using other (auxiliary) matrices. To this end we first neglect the $i \neq j$ conditions in~\eqref{eq:BBC2Fdef}. This `non-diagonal part' of the BBC2 functional can now be expressed in a separable form as
\begin{align}\label{eq:BBC2noDiag}
W^{\text{BBC2}}_{\text{no diag}}
&= \half\Trace\Bigl\{\mat{\gamma}\cdot\mat{v}^{\text{J}}(\mat{\gamma}) +
\sqrt{\mat{\gamma}}^{\text{virt}}\cdot\mat{v}^{\text{K}}\bigl(\sqrt{\mat{\gamma}}^{\text{virt}}\bigr) - {} \notag \\
&\eqspace\hphantom{\half\Trace\Bigl\{}
2\sqrt{\mat{\gamma}}^{\text{virt}}\cdot\mat{v}^{\text{K}}\bigl(\sqrt{\mat{\gamma}}^{\text{occ}}\bigr) -
\mat{\gamma}^{\text{occ}}\cdot\mat{v}^{\text{K}}\bigl(\mat{\gamma}^{\text{occ}}\bigr)\Bigr\} \, ,
\end{align}
where
\begin{subequations}
\begin{align}
f(\gamma)^{\text{occ}}_{\mu\nu} &\isDefinedAs
\sum_{\crampedclap{n_i \leq \nhalf}}\braket{\chi_{\mu}}{\phi_i}\, f(n_i) \, \braket{\phi_i}{\chi_{\nu}} \, , \\
f(\gamma)^{\text{virt}}_{\mu\nu} &\isDefinedAs
\sum_{\crampedclap{n_i > \nhalf}}\braket{\chi_{\mu}}{\phi_i}\, f(n_i) \, \braket{\phi_i}{\chi_{\nu}} \, .
\end{align}
\end{subequations}
The remaining diagonal part (correction) is now of the form
\begin{align}\label{eq:BBC2diag}
W^{\text{BBC2}}_{\text{diag}} = \half\sum_{\crampedclap{n_i \leq \nhalf}} (n_i^2 - n_i) \left[ ii | ii \right]  - 
\sum_{\crampedclap{n_i > \nhalf}} n_i \left[ ii | ii \right] \, .
\end{align}
Unfortunately the diagonal correction cannot be straightforwardly be written in a separable form and it seems that we still need to resort to a 4-index transformation of the two-electron integrals for their evaluation. However, the proposed trick can still be used by first constructing 1RDMs in which only one NO is occupied
\begin{align}
\bar{\gamma}^{(i)}_{\mu\nu} \isDefinedAs \braket{\chi_{\mu}}{\phi_i}\braket{\phi_i}{\chi_{\nu}} \, .
\end{align}
The next step is to form the contraction with the two-electron integrals as
\begin{align}
v^{(i)}_{\mu\nu} = \sum_{\kappa\lambda} \left[ \mu\nu | \kappa\lambda \right] \bar{\gamma}^{(i)}_{\lambda\kappa} \, .
\end{align}
The last step is to transform the potentials $\mat{v}^{(i)}$ back to the NO representation and to form the final contraction
\begin{align}\label{eq:diagEcontrib}
W^{\text{diag}}(\mat{d}) = \half\sum_id_iv^{(i)}_{ii} \, ,
\end{align}
where the elements $d_i$ depend on the particular form of the approximate 1RDM functional under consideration. For the BBC2 functional we have
\begin{align}
d_i = \begin{cases}
n_i^2 - n_i		&\text{for $n_i \geq \nhalf$} \\
-2n_i			&\text{for $n_i < \nhalf$} \, .
\end{cases}
\end{align}
The corresponding projected orbital derivatives required for the SCF can be obtained from the off-diagonal elements
\begin{align}\label{eq:diagDorb}
W^{\text{diag}}_{kl}(\mat{d}) &= d_kv^{(k)}_{kl} \, .
\end{align}
Though we could avoid the use of a 4-index transformation in this manner, the additional computational cost to calculate the diagonal correction of the BBC2 functional~\eqref{eq:BBC2diag} is significant compared to the cost to calculate the non-diagonal part~\eqref{eq:BBC2noDiag}. The non-diagonal part only needs the contraction with 4 different auxiliary matrices (1 Coulomb and 3 exchange), whereas the diagonal part requires a contraction with $m$ auxiliary matrices, so comprises a significantly more expensive part of the functional to evaluate. Furthermore, the complete construction of the orbital 1RMDs $\bar{\mat{\gamma}}^{(i)}$ and the corresponding potentials $\mat{v}^{(i)}$ gives a significant memory imprint, since both scale cubically with the number of basis functions. Fortunately, the special structure of the diagonal correction allows one to avoid the explicit construction of these large matrices. Avoiding the explicit construction of the orbital 1RDMs, $\bar{\mat{\gamma}}^{(i)}$, is readily achieved by not constructing them explicitly.
Instead, the required elements are only constructed on a need-to-be basis when looping over the two-electron integrals.

Now let us consider the high memory imprint of the potentials $\mat{v}^{(i)}$. Note that in the expression for the energy contribution~\eqref{eq:diagEcontrib} and for the orbital derivative~\eqref{eq:diagDorb}, at least one of the lower indices is always equal to the upper index, so we can avoid the construction of many unnecessary elements. We do this by transforming the potentials $\mat{v}^{(i)}$ partially to the NO basis immediately during their construction as
\begin{align}\label{eq:orbPotential}
\bar{v}_{i\nu} \isDefinedAs v^{(i)}_{i\nu} &= \sum_{\mu\kappa\lambda}
\braket{\phi_i}{\chi_{\mu}}\left[ \mu\nu | \kappa\lambda \right] \bar{\gamma}^{(i)}_{\lambda\kappa} \notag \\
&= \sum_{\mu\kappa\lambda}
\braket{\phi_i}{\chi_{\mu}}\left[ \mu\nu | \kappa\lambda \right] \braket{\chi_{\lambda}}{\phi_i}\braket{\phi_i}{\chi_{\kappa}} \, .
\end{align}
The projected orbital derivatives and the contribution to the energy can now easily be calculated by transforming the last index also to the NO basis and forming
\begin{subequations}
\begin{align}
W^{\text{diag}}_{kl}(\mat{d}) &= d_k\bar{v}_{kl} \, , \\
W^{\text{diag}}(\mat{d}) &= \half\sum_kd_k\bar{v}_{kk} = \half\Trace\bigl\{\mat{W}^{\text{diag}}\bigr\} \, .
\end{align}
\end{subequations}
Though we have now avoided the explicit construction of the $\bar{\mat{\gamma}}^{(i)}$ matrices and the corresponding potentials $\mat{v}^{(i)}$, the operation count for the calculation of the diagonal correction is significantly higher than for the non-diagonal part of the functional. For separable functionals with a diagonal correction, the evaluation of the diagonal part therefore remains the computational bottleneck in the evaluation of their values and derivatives.

The formal scaling of the proposed algorithm for functionals without diagonal corrections is still of order $m^4$ due to the loop over the two-electron integrals to construct the Coulomb potentials~\eqref{eq:HPotential} and exchange potentials~\eqref{eq:xPotential}. The situation is even worse for functionals with diagonal corrections, since the construction of the intermediate potentials $\bar{\mat{v}}$~\eqref{eq:orbPotential} makes it formally an $m^5$ process. The main advantage of the proposed algorithm is that integral screening becomes way more effective than in a calculation via a 4-index transformation. Screening of the two-electron integrals is a very common strategy in quantum chemistry software to avoid the calculation of many insignificant two-electron integrals. The number of significant two-electron integrals turns out to be only of the order $m^2$ for the larger molecular systems of interest,\cite{HelgakerJorgensenOlsen2000} so the scaling can in principle be reduced by an additional factor 2. This is readily done by using the Schwarz inequality on the two-electron integrals~\cite{Whitten1973}
\begin{align}
0 \leq \left[ ij | kl \right] \leq \sqrt{\left[ ij | ij \right]} \, \sqrt{\left[ kl | kl \right]} \, .
\end{align}
The integrals on the right-hand side are pre-calculated and only require $m^2$ storage. When the product of the square roots on the right-hand side are larger than some tolerance $\epsilon$, the program will actually calculate the integrals on the left. Screening of two-electron is particularly effective when the integrals are evaluated in a basis with a strong local character. This method and more advanced techniques have been implemented in virtually all quantum chemistry software packages, so can directly be exploited to reduce the scaling significantly as I will demonstrate in the following section.

It should be mentioned that the importance of separability for an efficient evaluation of energy expressions and derivatives has already been recognised for a few decades in quantum chemistry. The most famous example is the 2\textsuperscript{nd} order Møller--Plesset (MP2) energy correction, which in its usual form is non-separable
\begin{align}
E^{\text{MP2}} = -\frac{1}{4}\sum_{ijab} \frac{\abs[\big]{[a i | b j] - [a j | b i]}^2}{\epsilon_a + \epsilon_b - \epsilon_i - \epsilon_j} \, ,
\end{align}
where the indices $i,j$ refer to the occupied Hartree--Fock (HF) orbitals, the indices $a,b$ to the unoccupied HF orbitals and $\epsilon_r$ are the HF orbital energies. By writing the denominator as a Laplace transform, the MP2 energy correction can be turned into a separable formt\cite{Almlof1991, HaserAlmlof1992, Haser1993}
\begin{align}
E^{\text{MP2}} =-\frac{1}{4}\binteg{t}{0}{\infty}\sum_{ijab}\e^{-(\epsilon_a + \epsilon_b - \epsilon_i - \epsilon_j)t}\abs[\big]{[a i | b j] - [a j | b i]}^2 \, .
\end{align}
The summation over the HF orbitals is now a separable expression and the strategy described before can now be used to evaluate the integrant efficiently. The price to pay is that the contraction needs to be evaluated for several values of $t$ to evaluate the integral numerically with some suitable numerical integration scheme, but this only increases the computational cost by a prefactor. A similar trick can be used to bring the random phase approximation (RPA) in the linear scaling regime via the spherical Laplace transform.\cite{SchurkusOchsenfeld2016} Similar tricks will probably be useful to rewrite additional approximate 1RDM functionals in a separable form.

\section{Benchmarking}
\label{sec:benchmark}
To test the new strategy to evaluate the energy of separable functionals, I have constructed a modular Fortran 2003/2008 implementation, interfaced to a modified version of the \textsc{Gamess-US} program package.\cite{SchmidtBaldridgeBoatz1993, GordonSchmidt2005, gamess-us} The build-system of \textsc{Gamess-US} has been replaced by \textsc{foray},\cite{foray} to avoid figuring out module dependencies by hand and allow for parallel compilation. The implementation is currently only intended for serial runs. A parallel implementation will be considered later. The cut-off criterion for the two-electron integrals has been set to $\epsilon = 10^{-10}$.

\begin{figure}[t]
  \includegraphics[width=\columnwidth]{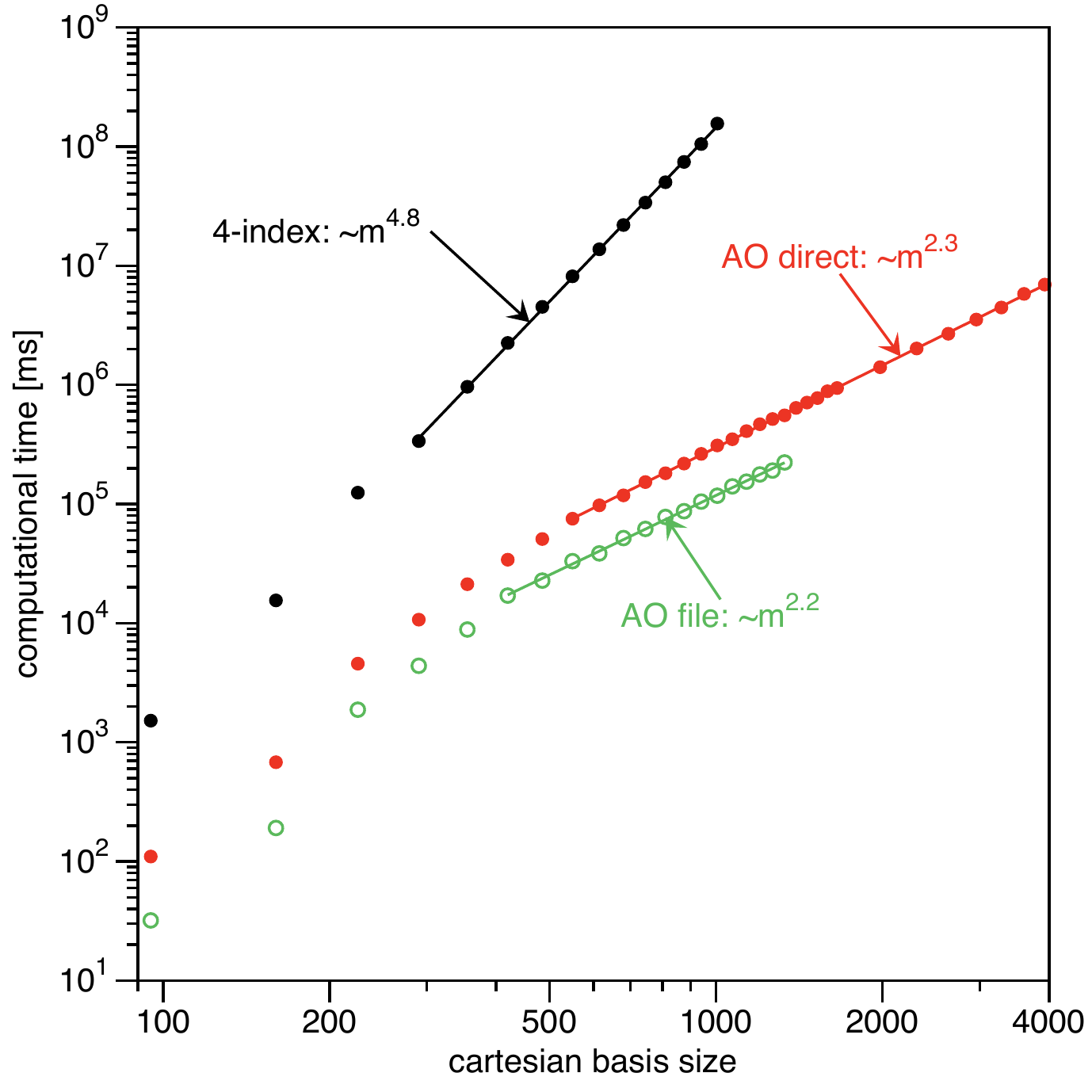}
  \caption{Log-log plot of the computational time for the evaluation of the energy and gradient from the Müller functional~\eqref{eq:Mueller} against the number of primitive basis functions in the calculation. The straight lines indicate which points have been included in the fits to extract the exponents.}
  \label{fig:Mueller}
\end{figure}

\begin{figure}[t]
  \includegraphics[width=\columnwidth]{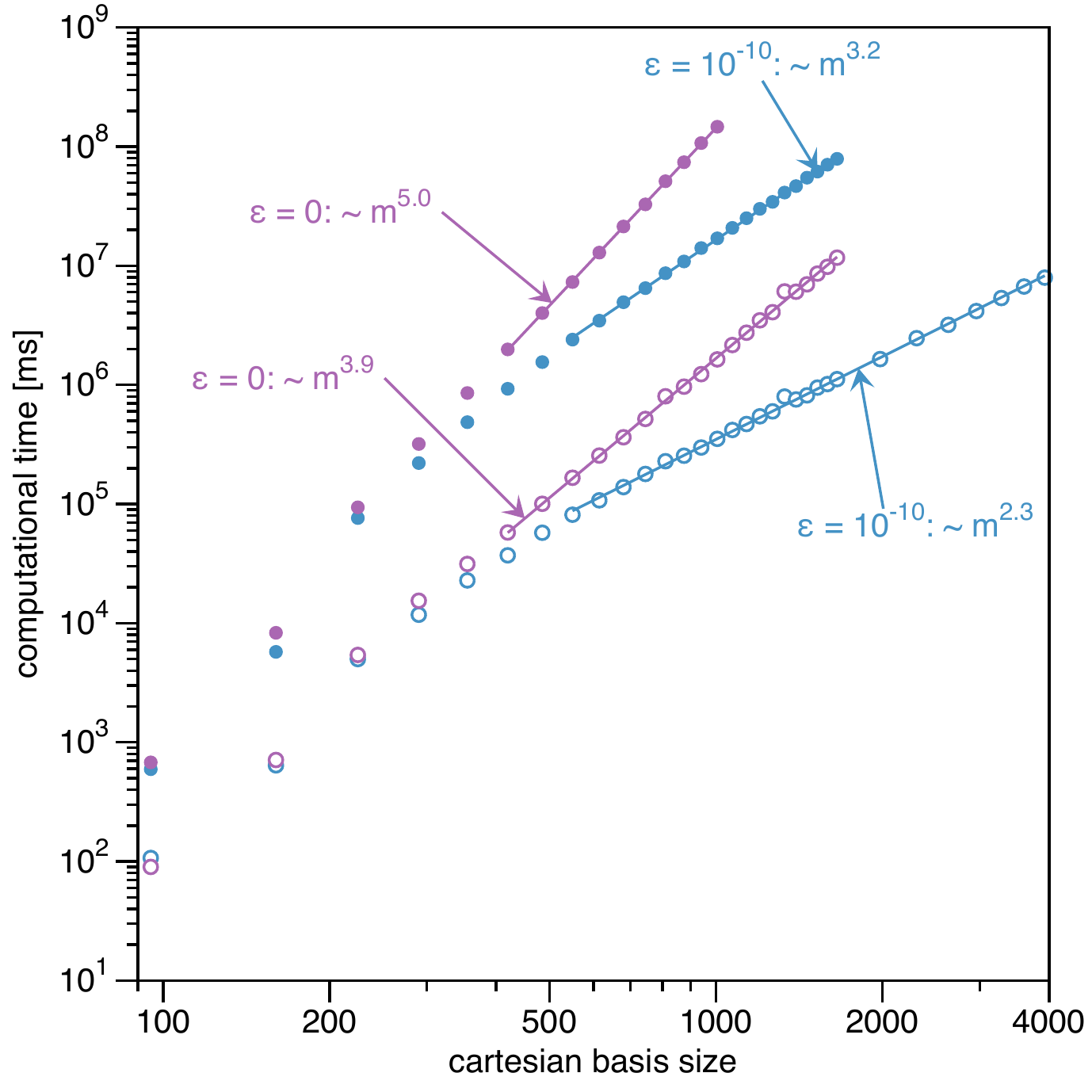}
  \caption{Log-log plot of the computational time for the evaluation of energy and gradient from the BBC2 functional. The filled circles correspond to the full BBC2 functional and the open circles to only the non-diagonal part~\eqref{eq:BBC2noDiag}. The straight lines indicate which points have been included in the fits to extract the exponents.}
  \label{fig:BBC2timing}
\end{figure}

The evaluation of the energy and gradients has been implemented in 3 different manners
\begin{enumerate}
\item The straightforward version using the 4-index transformation of the two-electron integrals.
\item Evaluation in AO basis using integrals stored on disk.
\item A direct option which does not store the two-electron integrals, but recalculates them each time when they are needed.
\end{enumerate}
In Fig.~\ref{fig:Mueller} I plot timings for the Müller functional~\eqref{eq:Mueller}. For the input 1RDM I have used the HF orbitals as NOs. Typically, the occupation numbers are fractional in 1RDM calculations. This has been mimicked to some extend by setting the occupations of the occupied HF orbitals to $0.9$ and the occupations of the virtual HF orbitals to $0.1\,N / (m - N)$. As test systems I used (linear) alkanes C$_n$H$_{2n+2}$ in a spherical cc-pVTZ basis.\cite{cc-pVT/QZ_H_B-Ne_aug_H} The highest point group symmetry of each system has been used to calculate only the symmetry unique integrals. For even $n$ the point group is $C_{2h}$ and for odd $n$ this is $C_{2v}$. The two smallest alkanes allow for the use of higher point groups ($T_d$ for methane and $D_{3d}$ for ethane). This additional reduction in symmetry unique two-electron integrals is reflected by the strong deviation from the general trend in the plot in Fig.~\ref{fig:Mueller}. The asymptotic scaling behavior (exponent) of the different strategies has been estimated by fitting a power function, $A\, \e^{\alpha\, m}$, to the results for a large enough basis size. The exact points taken into account varies per calculation and has been indicated by the straight line connection the points in Figs~\ref{fig:Mueller} and~\ref{fig:BBC2timing}.

The naïve implementation via the 4-index transformation of the two-electron integrals is clearly the most expensive one in Fig.~\ref{fig:Mueller}. Its main cost is the 4-index transformation, which is reflected in its high asymptotic scaling of order $m^{4.8}$. The evaluation in the AO basis is clearly more efficient. Reading the integrals from file is slightly faster than recalculating them, but the asymptotic scaling is basically the same. In this benchmark, the file reading is particularly fast due to the solid state drive (SSD) and the direct option is actually slow, since only one core is used. It is therefore expected that a parallel implementation would easily beat the `AO file' option, since the different processes would only start to compete for disk access. An additional bottle neck for the `AO file' option is that the two-electron integral file becomes excessively large. The last reported point for the `AO file' option (C$_{20}$H$_{42}$: 1330 cartesian basis functions) has already a two-electron file of 250 GB. The AO direct option does not have this bottle neck and can easily handle larger systems. The scaling of the `AO direct' option has been estimated to order $m^{2.3}$. The exponent has not converged however, since the formation of the required 1RDMs in AO basis ($\mat{\gamma}$ and $\sqrt{\mat{\gamma}}$) involves a matrix-matrix product which should scale worse. Assuming the BLAS implementation is based on Strassen's algorithm,\cite{Strassen1969} an asymptotic scaling of at least order $m^{\log_27} \approx m^{2.8}$ would be expected.

In Fig.~\ref{fig:BBC2timing} the computational time for the evaluation of the BBC2 functional with and without its diagonal part~\eqref{eq:BBC2diag} are shown when using the `AO direct' option. The first thing to notice is the effectiveness of screening by the Schwarz inequality. Both for the full BBC2 functional and only the non-diagonal part, the use of screening leads to a drastic reduction in the asymptotic scaling.

\begin{figure}[t]
  \includegraphics[width=\columnwidth]{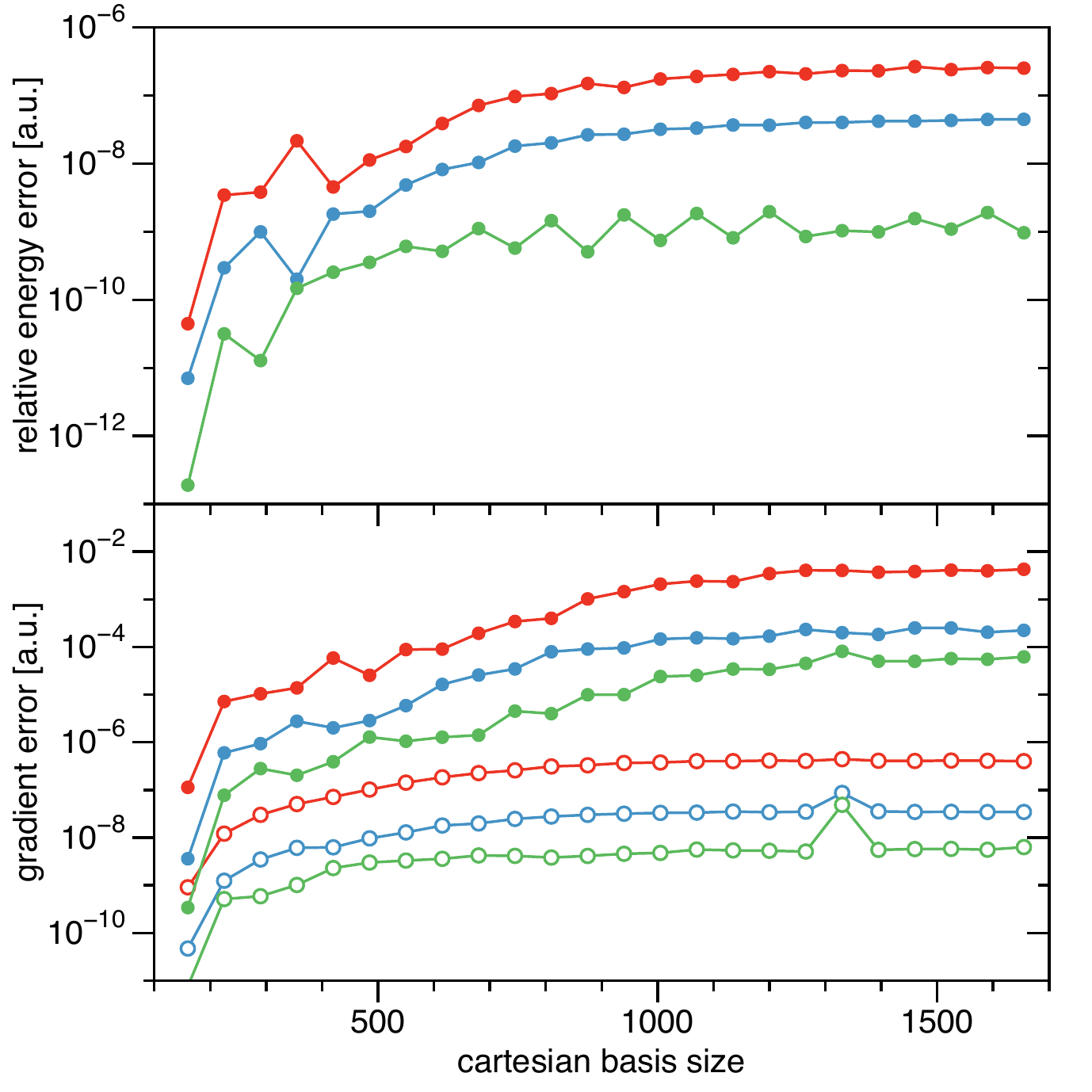}
  \caption{Plot of the relative absolute errors in the calculation of the energy (upper panel) and gradient (lower panel) of the non-diagonal part of the BBC2 functional for different cutoff criteria: upper curve (red) $\epsilon = 10^{-9}$, middle curve (blue) $\epsilon = 10^{-10}$, lower curve (green) $\epsilon = 10^{-11}$. The full circles in the gradient panel are the maximum errors in the gradient and the open circles are the averaged absolute error in the gradient.}
  \label{fig:BBC2noDiagError}
\end{figure}

Now let us concentrate on the calculations with only the non-diagonal part of the BBC2 functional~\eqref{eq:BBC2noDiag}. Though the non-diagonal part of the BBC2 functional requires the contraction with 4 different 1RDMs, instead of only 2 as required for the Müller functional, the increase in computational cost is marginal. Only the prefactor is increased slightly by $\bigl(A_{\text{BBC2}} - A_{\text{Müller}}\bigr)/A_{\text{Müller}} = 5\%$, but the asymptotic scaling remains the same, $m^{2.3}$.

On the other hand, the calculation of the diagonal corrections poses a significant increase in computational cost. They add an order of magnitude to the computational cost. If no screening is used, this even means that the asymptotic scaling is similar to the scaling of the 4-index transformation. Avoiding the use of small integrals is therefore crucial for the algorithm to improve over the standard evaluation by the 4-index transformation. If screening is used, however, the use of an AO based evaluation of the diagonal correction provides a major increase in efficiency compared to the traditional approach via a 4-index transformation. The asymptotic computational cost has been reduced from order $m^{5}$ to $m^{3.2}$, so much larger quantum systems are accessible with the new implementation, especially when the algorithm is additionally parallelized.

\begin{figure}[t]
  \includegraphics[width=\columnwidth]{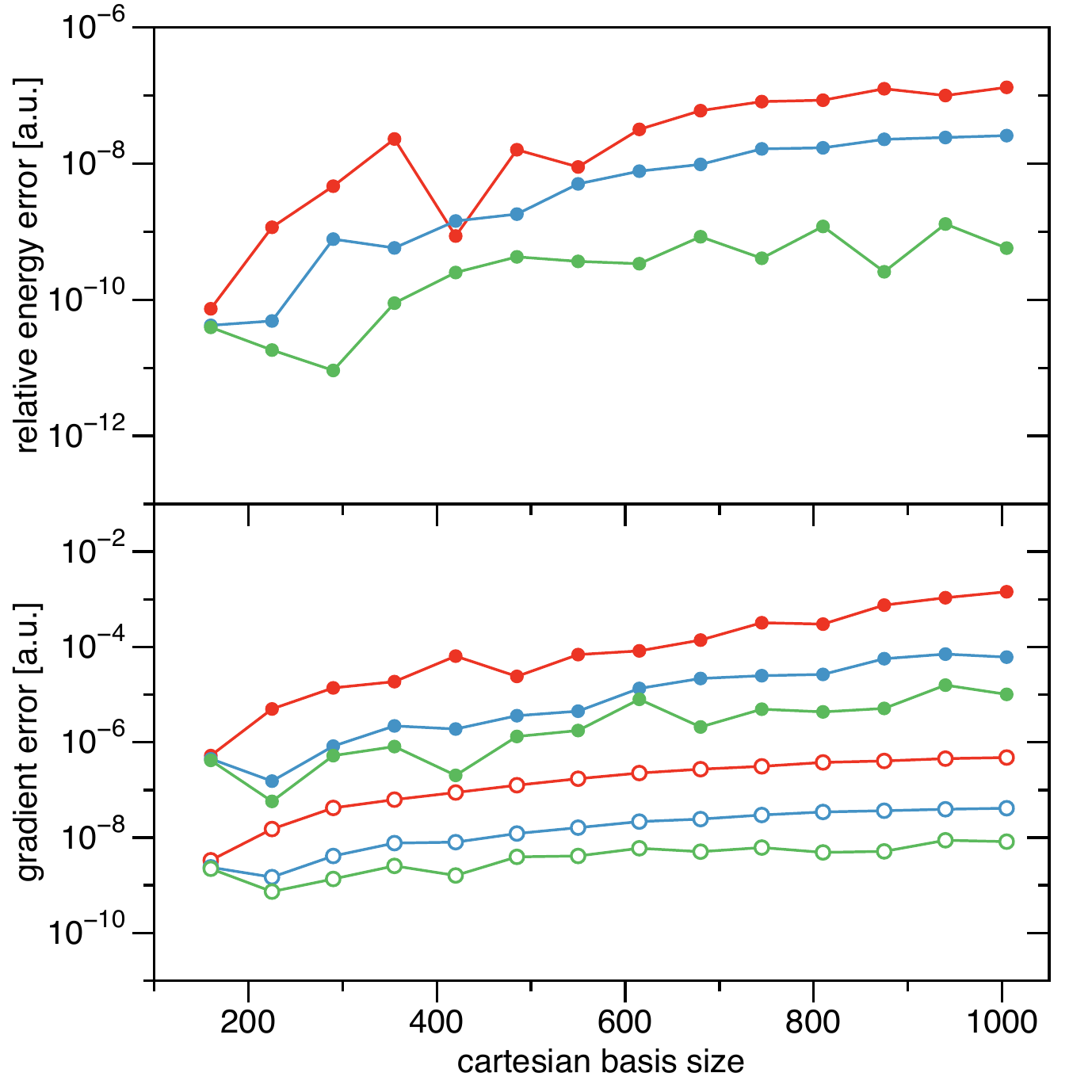}
  \caption{Similar as in Fig.~\ref{fig:BBC2noDiagError}, though now for the full BBC2 functional. Since the full BBC2 functional is much more expensive, only calculations up to C$_{15}$H$_{32}$ have been included in the comparison.}
  \label{fig:BBC2fullError}
\end{figure}

The effect of the cutoff criterion on the total energies and gradients has also been investigated. In the upper panel of Fig.~\ref{fig:BBC2noDiagError} the relative absolute error in the energy, $\abs{(E_{\epsilon} - E_0) / E_0}$, has been plotted as a function of the cartesian basis size for the alkane series. For the full range of alkanes, the error in the energy can be systematically reduced by lowering the value of the cutoff parameter, $\epsilon$. Only for the small alkanes the reduction in the error is more erratic, probably due to more or less error cancelations. In the lower panel of Fig.~\ref{fig:BBC2noDiagError} the averaged absolute error (open circles) and maximum error (filled circles) is shown for the alkane series. Again, the error in the gradient can be systematically reduced by lowering the cutoff parameter, $\epsilon$.

In Fig.~\ref{fig:BBC2fullError} I show the same errors for the full BBC2 functional for the different cutoff parameters, $\epsilon = \{10^{-9}, 10^{-10}, 10^{-11}\}$. We see that the inclusion of the diagonal contributions to the BBC2 functional~\eqref{eq:BBC2diag} looks very similar for the larger alkanes, So the error in the energy and gradient in the same ballpark when the diagonal part of the BBC2 functional is included. Only for smallest alkanes the errors are somewhat larger, probably due to less fortuitous error cancellations.

\section{Conclusion}
\label{sec:conclusion}
An algorithm to avoid the 4-index transformation for non-trivial separable 1RDM functionals has been presented, which reduces the formal computational cost to $m^4$. Diagonal corrections could be handled with a similar strategy, though additional modifications were needed to avoid excessive use of memory. The formal scaling of the diagonal corrections remains $m^5$ unfortunately. The main advantage of the newly proposed strategy for the evaluation of 1RDM functionals and their derivatives, is that now integral screening techniques can be used more effectively, leading to a significant reduction in the asymptotic scaling of the computational cost. Using screening by the Schwarz inequality, I have shown that the asymptotic scaling can be reduced to $m^{3.2}$ for separable functionals including diagonal corrections. For separable functionals without diagonal corrections the asymptotic scaling is even further reduced to $m^{2.3}$. It is expected, however, that for larger systems the scaling for functionals without diagonal corrections should be about order $m^{2.8}$ due to the matrix-matrix products involved. All in all, a significant boost in the speed of the evaluation of approximate 1RDM functionals has been achieved. As the evaluation was one of the major computational bottlenecks in practical 1RDM functional calculations. This is an important step to make the use of 1RDM functionals feasible for large molecular systems.

\section{Acknowledgements}
I am happy to dedicate this paper to Evert Jan Baerends, who has been a great Ph.D. supervisor. I especially appreciate his confidence and trust that I would find my own way and to develop my interests. Our discussions have always been very helpful to me to structure and organize my ideas well beyond my Ph.D.

Support from the Netherlands Foundation for Research NWO (722.012.013) through a VENI grant is gratefully acknowledged.

\bibliography{bible} 
\bibliographystyle{rsc} 

\end{document}